\documentclass[showpacs,prb,amsmath,amssymb]{revtex4}
\usepackage{graphicx}

\begin{document}

\title{The Superconductivity, Intragrain Penetration Depth and Meissner
Effect of RuSr$_{2}$(Gd,Ce)$_{2}$Cu$_{2}$O$_{10+\delta }$\\
}
\author{Y. Y. Xue}
\author{B. Lorenz}
\author{A. Baikalov}
\author{D. H. Cao}
\author{Z. G. Li}
\author{C. W. Chu}
\altaffiliation[Also at ]{Lawrence Berkeley National Laboratory, 1 Cyclotron Road, Berkeley, CA 94720;
and Hong Kong University of Science and Technology, Hong Kong}
\affiliation{Physics Department, and Texas Center for Superconductivity, University of
Houston, Houston TX 77204-5002\\
}
\date{\today }

\begin{abstract}
The hole concentration \textit{p }($\delta $), the transition temperature T$%
_{c}$, the intragrain penetration depth $\lambda $,\ and the Meissner effect
were measured for annealed RuSr$_{2}$(Gd,Ce)$_{2}$Cu$_{2}$O$_{10+\delta }$
samples. The intragrain superconducting transition temperature T$_{c}$
varied from 17 to 40 K while the \textit{p} changed by only 0.03 holes/CuO$%
_{2}$. The intragrain superfluid-density 1/$\lambda ^{2}$ and the
diamagnetic drop of the field-cooled magnetization across T$_{c}$ (the
Meissner effect), however, increased more than 10 times. All of these
findings are in disagreement with both the T$_{c}$ vs. \textit{p} and the T$%
_{c}$ vs. 1/$\lambda ^{2}$ correlations proposed for homogeneous cuprates,
but are in line with a possible phase-separation and the granularity
associated with it.
\end{abstract}

\maketitle




One key question about RuSr$_{2}$GdCu$_{2}$O$_{8+\delta }$ (Ru1212) and RuSr$%
_{2}$(Gd,Ce)$_{2}$Cu$_{2}$O$_{10+\delta }$ (Ru1222), where the partially
ferromagnetically (FM) aligned Ru-spins and the superconducting (SC) CuO$%
_{2}$ layers are structurally adjacent, is what determines their
superconductivity. Different groups have emphasized either their underdoped
nature (\textit{i.e.} the hole-concentration $p << 0.16$~holes/CuO$_{2}$) or 
the competition between the SC and
the coexisting FM through the Fulde-Ferrell-Larkin-Ovchinnikov (FFLO)\ phase.%
\cite{ber,chu,ful,pic,zhu} A verification was difficult until the recent
reports that the transition temperature T$_{c}$'s of Ru1222 and Ru1212 can
be adjusted over a broad range by oxygen-annealing and Cu-substitution,
respectively.\cite{bau,fel,cla} The reported data, unfortunately, show that
the T$_{c}$-enhancement is accompanied by both a \textit{p}-increase and a
suppression of the ferromagnetic spin-order, and offer no clear distinction.
To explore the topic, we measured the intragrain T$_{c}$, the \textit{p},
the intragrain superfluid-density 1/$\lambda ^{2}$, and the Meissner effect
in several annealed Ru1222 samples. The data were then compared with both
the T$_{c}$ vs. \textit{p} correlation,\cite{pre} and the T$_{c}$ vs. 1/$%
\lambda ^{2}$ line proposed.\cite{nac} We observed that a two-fold
enhancement of T$_{c}$ (from 17 to 40 K) is accompanied by a relatively
small change of \textit{p }(from 0.09 to 0.12 holes/CuO$_{2}$), but a
20-fold increase in 1/$\lambda ^{2}$ (from 0.3 to 6 $\mu $m$^{-2}$).
Together with an extremely large field effect of dT$_{c}$/dH $>$ 100 K/T and
a linear increase of the Meissner fraction with 1/$\lambda ^{2}$, the data
suggest that those ruthenocuprate grains are actually
Josephson-junction-arrays (JJA), in agreement with the phase-separation
model suggested.\cite{xuea} Further investigations of local magnetic
structures are needed to solve the problem.

Ceramic RuSr$_{2}$(Gd$_{0.7}$Ce$_{0.3}$)$_{2}$Cu$_{2}$O$_{10+\delta }$
samples were synthesized following the standard solid-state-reaction
procedure. Precursors were first prepared by calcined commercial oxides at
400-900 $^{\text{o}}$C under flowing O$_{2}$. Mixed powder with a proper
cation ratio was then pressed into pellets and sintered at 900 $^{\text{o}}$%
C in air for 24 hr. The final heat treatment of the ceramics was done at
1090 $^{\text{o}}$C for 60 hr after repeatedly sintering and regrinding at
1000 $^{\text{o}}$C. Powders with different particle sizes were prepared
according to the procedures previously reported.\cite{xue} The structure of
the samples was determined by powder X-ray diffraction (XRD), using a Rigaku
DMAX-IIIB diffractometer. The XRD pattern of a typical sample is shown in
Fig. 1. Refinement was done based on a space group of I4/mmm with lattice
parameters of \textit{a} = 3.839(1) and \textit{c} = 28.591(5) using the
Rietan-2000 program.\cite{rie} There are no noticeable impurity lines in the
X-ray diffraction pattern within our resolution of a few percent. The grain
sizes ($\approx $ 2-20 $\mu $m) of the ceramic samples, as well as the
particle sizes of the powders, were measured using a JEOL JSM 6400 scanning
electron microscope (SEM). The magnetizations were measured using a Quantum
Design SQUID magnetometer with an \textit{ac} attachment.

\begin{figure*}
\includegraphics[scale=.5]{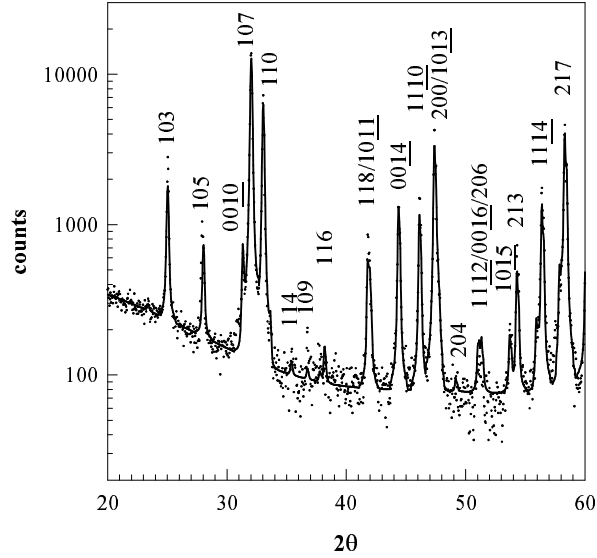}
\caption{\label{fig:fig1}The XRD of sample A0. dots:
data; solid line: the refinement with I4/mmm.}
\end{figure*}

\begin{figure*}
\includegraphics[scale=.5]{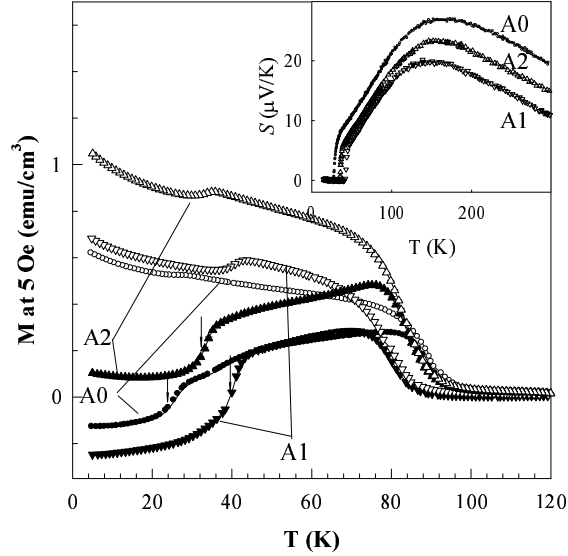}
\caption{\label{fig:fig2}The M$_{ZFC}$ (solid
symbols) and M$_{FC}$ (open symbols) at 5 Oe for $\bigcirc /\bullet $:
sample A0; $\bigtriangledown /\blacktriangledown $: sample A1; and $%
\bigtriangleup /\blacktriangle $: sample A2. Inset: Thermal power \textit{S}%
(T) of the three samples.}
\end{figure*}

The data from nine Ru1222 samples in three sets are presented here. A0, B0,
and C0 are as-synthesized samples with slightly different superconducting
transition temperatures between 26 and 30 K (probably due to the slight
differences in the final heat treatment and the various resulting intragrain
granularities). All others are pieces of the respective as-synthesized
ceramic after a 2 hr/600 $^{\text{o}}$C anneal. The gases used in annealing
are 300 atm. O$_{2}$ for samples A1, B1, and C1; 20 atm. O$_{2}$ for sample
A2; Ar+0.01 atm. O$_{2}$ for sample C2; and high purity Ar (99.99\%) for
sample C3. The zero-field-cooled magnetization (M$_{ZFC}$) and the
field-cooled one (M$_{FC}$) of samples A0, A1, and A2 are shown in Fig. 2. A
systematic increase of the T$_{c}$\ (defined as the major inflection point
of the M$_{ZFC}$ and marked by arrows in the figure) with the assumed
oxygen-intake can be clearly seen, \textit{i.e.} from 26 to 40 K.\cite{fel}
It should be pointed out that the T$_{c}$ so-defined is actually the
intragrain transition-temperature based on both the size- and the H$_{ac}$%
-dependencies of the \textit{ac} $\chi $ measured in the same samples, where
H$_{ac}$ is the \textit{ac} field used (Fig. 3).\cite{xue,lor} The
intergrain transition of our Ru1222 samples is typically $\approx $ 10-20 K
lower, and a prefect shielding can be reached only under H$_{ac}$ $\approx $
0.01 Oe (Fig. 3). This relatively weaker intergrain coupling of Ru1222
ceramic seems to be typical in previously published reports, where a less
than 100\% shielding in M$_{ZFC}$ was observed.\cite{fel,wil} We attribute
this to the 1090 $^{\text{o}}$C final heat-treatment temperature used, which
is higher than that used for Ru1212.

To estimate the \textit{p}, the thermoelectric power, \textit{S}, was
measured (inset, Fig. 2). The overall T-dependence of the \textit{S} is
similar to that of the underdoped YBa$_{2}$Ca$_{3}$O$_{7-\delta }$, and is
in agreement with the data previously reported.\cite{wil} No evidence of the
RuO$_{2}$ contributions can be noticed.\cite{ber} A moderate increase of 
\textit{p} (\textit{i.e.} from 0.104 holes/CuO$_{2}$ in sample A0 to 0.121
holes/CuO$_{2}$ in sample A1) was then deduced using the proposed universal
correlation of \textit{S}(290 K) = 992exp(-38.1\textit{p}).\cite{obe} To
verify the deduced \textit{p},\cite{wil} the oxygen-intake, $\Delta \delta $%
, was measured using a gas-effusion cell, where the sample was heated to 800 
$^{\text{o}}$C and the released oxygen was measured by both a pressure-gauge
and a mass spectrometer.\cite{ham} The total oxygen released is 0.1 and
0.115 O/Ru1222 for samples A0 and A1, respectively. While the absolute
stoichiometry, 10+$\delta $, may depend on the phase compositions at 800 $^{%
\text{o}}$C, the $\Delta \delta $ should be less debatable. The $\Delta
\delta $ $\approx $ 0.015 so obtained is in good agreement with both the
estimated \textit{p} from \textit{S }and the\ reported rate of $\Delta
\delta $/$\Delta T_{c}\approx $ 0.0014 O/K for Ru1222 with similar T$_{c}$.%
\cite{otz,note} The expected T$_{c}$ enhancement, however, should be less
than 6 K based on the correlation of T$_{c}$ = T$_{c,\max }$[1-82(\textit{p}%
-0.16)$^{2}$] with T$_{c,\max }$ $\leq $ 50 K.\cite{obe} This value is far
smaller than the 14 K enhancement observed, and the change in carrier
concentration should not be the dominant factor.

\begin{figure*}
\includegraphics[scale=.5]{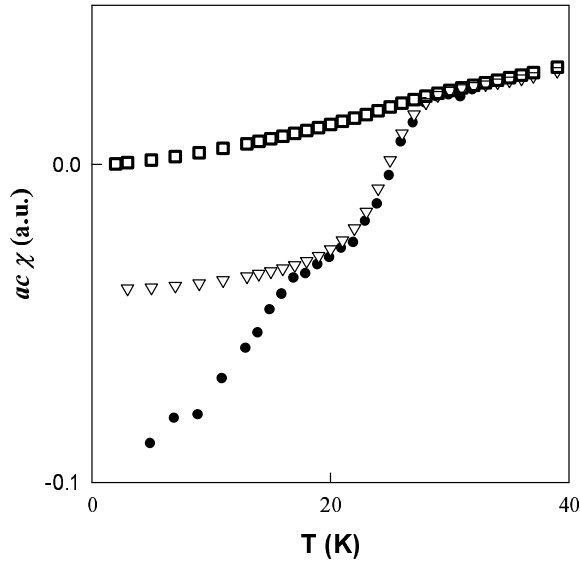}
\caption{\label{fig:fig3}$ac\chi$ of
the bulk sample A0 (powders with particle size of 10 $\mu$m or
larger show the same $\chi$) under H$_{ac}$ of 3 ($\bigtriangledown 
$) and 0.01 Oe ($\bullet $), as well as that at H$_{ac}$ = 3 Oe of a 1 $\mu$m 
powder made from it ($\square $).}
\end{figure*}

It is interesting to note that the T$_{c}$-enhancement is accompanied by an
increase in the diamagnetic drops ($\Delta $M$_{ZFC}$, $\Delta $M$_{FC}$)
across T$_{c}$ in both M$_{ZFC}$ and M$_{FC}$ . For example, the $\Delta $M$%
_{ZFC}$ is $\approx $ 0.15, 0.2, and 0.3 emu/cm$^{3}$ at 5 Oe for samples
A0, A2, and A1, respectively (Fig. 2).\cite{note2} Comparable trends can
also be noticed in the data previous reported.\cite{fel} We tentatively
attribute the change in $\Delta $M$_{ZFC}$ to a decrease in 1/$\lambda ^{2}$
(actually 1/$\lambda _{ab}^{2}$), and verify the interpretation by the
direct measurement of the intragrain penetration-depths $\lambda $ through
the \textit{ac} susceptibility of powders. Several powders were prepared
from the same ceramic by sorted according to their particle sizes, and the $%
\lambda $ was deduced from the size dependency of the $\chi $ observed. The
details have been reported before.\cite{che,xue} The previous data analysis
procedure, however, was slightly modified here to simultaneously fit both
the large magnetic background $\chi _{m}$ and $\lambda $. For a
randomly-oriented power \textit{j} (=1...n), which contains particles $%
i=1...m$ with sizes of $d_{j,i}$, one has:

$\chi _{j}=\sum_{i}\int \{[1-6(\lambda _{ab}/d_{j,i})coth(d_{j,i}/2\lambda
_{ab})+12(\lambda _{ab}/d_{j,i})^{2}]d_{j,i}^{3}cos^{2}\theta sin\theta $

$+[1-6(\lambda _{c}/d_{j,i})coth(d_{j,i}/2\lambda _{c})+12(\lambda
_{c}/d_{j,i})^{2}]d_{j,i}^{3}sin^{2}\theta sin\theta \}d\theta
/\sum_{i}d_{j,i}^{3}+\chi _{m}$, where $\lambda _{c},$ $\lambda _{ab}$, and $%
\theta $ are the penetration depths along \textit{c}, \textit{ab}, and the
polar angle, respectively. When $\lambda _{c}$ $>>$ $\lambda _{ab}$ (highly
anisotropy approximation) one has:

$\chi _{j}=$ $\sum_{i}\{[1-6(\lambda _{ab}/d_{j,i})coth(d_{j,i}/2\lambda
_{ab})+12(\lambda _{ab}/d_{j,i})^{2}]d_{j,i}^{3}/3\sum_{i}d_{j,i}^{3}+\chi
_{m}$.

A regression was used to calculate $\lambda $ without assumptions of $\chi
_{m}$. The $\chi _{j}$ of the powder with the smallest average particle size
was used as the initial value of $\chi _{m}$. The initial value of $\lambda $
was deduced from the $\chi $ of the powder with the largest particle size.
The new $\chi _{m}$ was then regressively calculated through a least-square
fit using the approximate $\lambda $ value and the $\chi _{j}$, $d_{j,i}$
data observed. The regression of $\lambda $ followed. Typically, the result
will be convergent to within 1\% after three regression cycles.

\begin{figure*}
\includegraphics[scale=.5]{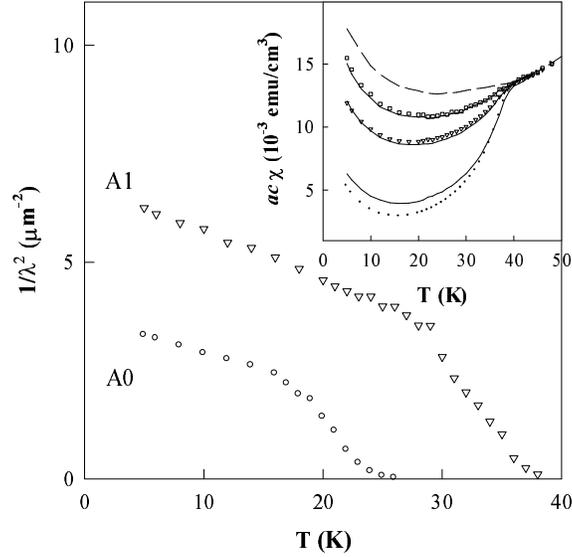}
\caption{\label{fig:fig4}The intragrain 1/$\lambda ^{2}$ for $\bigtriangledown $: sample A0 and 
$\bigcirc $: sample A1.
Inset: the \textit{ac }$\chi $ of powder made from sample A1 with
particle size of $\bigcirc $: 1.9 $\mu $m; $\bigtriangledown $: 1.3 $\mu $m; 
and $\square $: 0.9 $\mu $m. The solid lines are fits
and the dashed line is the estimated magnetic background.}
\end{figure*}

The procedure was tested on the data of YBa$_{2}$Cu$_{3}$O$_{6.6}$ and MgB$%
_{2}$ previously collected, and led to a good agreement with the values
expected.\cite{xue,che} The possible uncertainty of the $\lambda $ seems to
be less than 30\%, mainly from the statistical uncertainty of $d_{i}$. As an
example, the data of three powder-samples from ceramic A1 with effective
particle-sizes of \textit{d} = 1.9, 1.3, and 0.9 $\mu $m are shown in the
inset of Fig. 4. Their $\chi $, the deduced $\chi _{m}$, and the fitting
results are shown as symbols, a dashed line, and solid lines, respectively.
The deduced superfluid densities, 1/$\lambda ^{2}$'s of samples A0 and A1
are shown in Fig. 4. The twofold increase of 1/$\lambda ^{2}$, which is in
agreement with the raw M$_{ZFC}$ data of both that in the Fig. 2 and that
reported previously,\cite{fel,cla} confirms the above assumption that the
change in 1/$\lambda ^{2}$ is the dominant factor for $\Delta $M$_{ZFC}$ and 
$\Delta $M$_{FC}$. This view is further supported by an unusual large 
\textit{d}T$_{c}$/\textit{d}H $\approx $ 100 K/T similar to that observed in
Ru1212Eu (Fig. 5).

\begin{figure*}
\includegraphics[scale=.5]{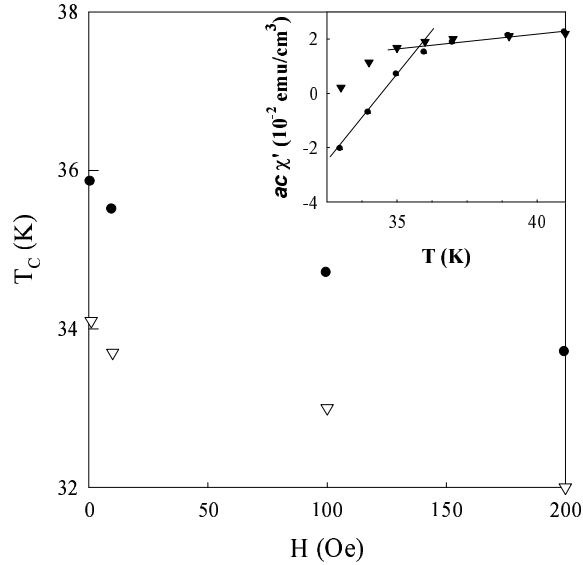}
\caption{\label{fig:fig5}The
intragrain T$_{c}$, deduced as the onset of the differential $\chi $
($\bullet $) and the inflection point of M$_{FC}$ ($\bigtriangledown $).
Inset. The differential susceptibility measured as the \textit{ac} $\chi $ 
at H$_{ac}$ = 0.3 Oe with a \textit{dc} bias of 1 ($\bullet $) and
100 Oe ($\blacktriangledown $). The T$_{c}$ is determined as the cross point
of the linear fits (the solid lines) below and above T$_{c}$.}
\end{figure*}

The increase in 1/$\lambda ^{2}$\textit{\ }is significantly higher than that
expected from both the \textit{p} and the T$_{c}$ observed. In principle,
the effective mass \textit{m}*, pair-breaking scatterings, and intragrain
granularity can all affect the 1/$\lambda ^{2}$ observed. However, the
comparable 300 K resistivity of the samples A0 and A1, \textit{e.g.} $%
\approx $ 0.024 and 0.029 $\Omega $cm respectively, suggests that \textit{m}%
* may not play a major role here. To verify the possibility of the simple
pair-broken mechanism, the 1/$\lambda ^{2}$'s of several samples with T$_{c}$
ranging from 17 K (sample C3) to 40 K (sample A1) were measured (Fig. 6).%
\cite{note3} In typical cuprates with pair-breakers (\textit{e.g.} Zn),
Nachumi \textit{et. al}. have observed that T$_{c}$ is a universal linear
function of 1/$\lambda ^{2}$.\cite{nac} Although there are still disputes
about the data-details at high Zn-levels and their interpretation, all
reported T$_{c}$ data approach zero with the suppression of 1/$\lambda ^{2}$
with pair-breaking. In particular, both the strong-coupling \textit{d}-wave
model in the unitarity limit and the \textquotedblleft swiss
cheese\textquotedblright\ model predicts T$_{c}$ $<<$
10 K when 1/$\lambda ^{2}$ $\approx $ 0.3 $\mu $m$^{-2}$ (\textit{i.e.} $%
\sigma $ $\approx $ 0.02 $\mu $s$^{-1}$ in a $\mu $SR measurement).\cite%
{bera,nac}\ It is, therefore, interesting to note that all of our data
points fall on the far left of this line. In particular, the T$_{c}$ of
sample C2 with 1/$\lambda ^{2}$ $\approx $ 0.3 $\mu $m$^{-2}$ is still 15 K
or higher. The trend is in line with the data of Ru1212, where samples with $%
\lambda $(5 K) as large as 2-3 $\mu $m still have T$_{c}$ $>$ 20
K.\cite{xue} A simple pair-broken mechanism, therefore, may have difficulty
in accommodating the data. This view is supported by the fact that there is
no systematic correlation between the T$_{c}$ (or 1/$\lambda ^{2}$) and the M%
$_{FC}$ (or the remanent moment) above T$_{c}$, \textit{i.e.} the FM aligned
spins. In fact, the M$_{FC}$ at 5 Oe and 50 K differs less than 5\% between
samples C3 and C1 with T$_{c}$ of 17 and 37 K, respectively. The M$_{FC}$
spread for samples A0, A1, and A2 is slightly larger, but again without a
systematic dependency. A similar trend has also been pointed out previously
in both the Ce-doped Ru1222 and the Cu-substituted Ru1212.\cite{wil,cla}
Intragrain granularity, therefore, seems to be a more reasonable
interpretation. It was proposed that the T$_{c}$ of a JJA is 2.25\textit{J},%
\textit{\ }with \textit{J} being the coupling energy of a junction.\cite{li}
The $\lambda $, in such a case, may depend on the length of the junctions
involved, but the T$_{c}$ will not. A non-zero phase-lock temperature,
therefore, may coexist with an unusually long $\lambda $ if the junction
length is large.

Granularity of Ru1212 has been previously reported and attributed to either
structural defects or possible phase-separation.\cite{ber,chu,xue} However,
no correlations between the T$_{c}$ and the proposed 90$^{\text{o}}$-rotated 
\textit{a-c }microdomains (or the coherent rotation of the RuO$_{6}$
octahedrons) have ever been observed.\cite{mac,ber} We, therefore, favor a
mesoscopic phase-separation between FM and AFM species as the origin of the
granularity.\cite{xuea,had} The observation of a correlation between the
magnetic transition temperature and the granularity (T$_{c}$, 1/$\lambda ^{2}
$...) here certainly supports this view (Fig. 2). However, we will not
discuss it further since it is not essential for the topics concerned, 
\textit{i.e.} the evolution of T$_{c}$ and Meissner effect with granularity.%

\begin{figure*}
\includegraphics[scale=.5]{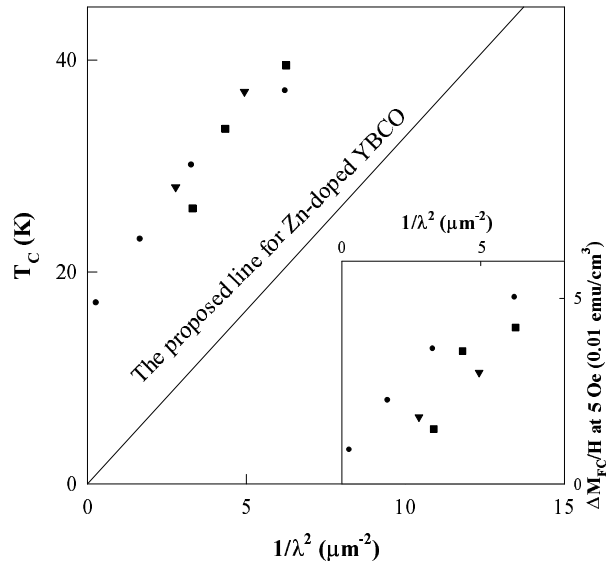}
\caption{\label{fig:fig6}T$_{c}$ vs. 1/$\lambda^{2}$ at 5 K for several annealed samples. Solid 
line: the Uemura line.
Inset: the diamagnetic drop $\Delta $M$_{FC}$/H across T$_{c}$ at 5 Oe vs. 
1/$\lambda ^{2}$. Symbols used are $\blacksquare $: A0-A2; $\blacktriangledown $:
B0-B1; $\bullet $: C0-C3.}
\end{figure*}

The systematic increase of $\Delta $M$_{FC}$, the Meissner effect, with
annealing is also obvious in Fig. 2. The large Meissner fraction in a Ru1212
sample below 1 Oe and its disappear once above 10 Oe has previously been
taken as evidence for a homogeneous superconductivity and a spontaneous
vortex state (SVS), respectively.\cite{ber} The interpretation, in our
opinion, is neither the only possible one nor the most likely one. Within
the framework of the Ginzburg-Landau (G-L) model, the reversible part of $%
\Delta $M$_{FC}$ of a type II superconductor should have a maximum of H$_{c1}
$/4$\pi $ at H$_{c1}$; and a few times smaller ($\approx $ $\frac{\phi _{%
\text{o}}\mathit{ln}(H_{c2}/H)}{32\pi ^{2}\lambda ^{2}}$ = $\frac{H_{c1}%
\mathit{ln}(H_{c2}/H)}{8\pi \ln \kappa }$) in mixed states far above H$_{c1}$%
, where H$_{c1}$, H$_{c2}$, and $\kappa $ are the lower- and upper-critical
fields and the G-L parameter, respectively.\cite{hao,note1} This is simply
the result of a competition between the magnetic energy M$\cdot $H and the
carrier kinetic-energy, and should hold even in the existence of spontaneous
vortex and an internal magnetic field B$_{M}$ (= 4$\pi $M in homogeneous\
ferromagnetic superconductors). In particular, the $\Delta $M$_{FC}$ should
be $\approx $ $\phi _{\text{o}}\ln \kappa $/(16$\pi ^{2}\lambda ^{2}$) ($>$
1 emu/cm$^{3}$ with $\lambda < 0.5$~$\mu$m) over a broad
H-range regardless of the value of B$_{M}$ if the pinning is weak. The
maximum $\Delta $M$_{FC}$, we would argue, is a far better parameter for
ferromagnetic superconductors than the widely used $\chi $, whose
interpretation may be ambiguous due to the uncertainties in the B$_{M}$ and
the possible SVS. The $\Delta $M$_{FC}$ of sample A1, for example, is $%
\approx $ 0.3 emu/cm$^{3}$ at 20 Oe, and seems to increase continuously with
H although the large magnetic background makes a quantitative estimation
difficult at larger fields. This value is not too far from that expected
based on the deduced $\lambda $ $\approx $ 0.4 $\mu $m, considering the
corrections needed for the random grain-orientations and vortex pinning. The
much smaller $\Delta $M$_{FC} < 0.04$~emu/cm$^{3}$ over the
whole H range reported) in Ref. 1, on the other hand, may imply a unusually
long $\lambda $, \textit{i.e.} severe granularity, if the pinning is not too
strong.

It should be noted that a 100\% Meissner effect can be reached in a JJA-like
heterogeneous superconductor below its effective H$_{c1}\propto $ 1/$\lambda
^{2}$. The value of $\Delta $M$_{FC}$/H is determined by both a free-energy
balance and the vertex pinning. For the energy balance, all of the
conclusions of Hao \textit{et al.} should still hold if the JJA parameters
are used.\cite{hao} In particular, a full Meissner effect can be expected
below $\phi _{\text{o}}$/(32$\pi ^{2}\lambda ^{2}$) $\approx $ 1 Oe with a $%
\lambda $ as large as 2 $\mu $m. The pinning strength, on the other hand,
depends only on these parameters averaged over a length-scale of vortex
cores. The dT$_{c}$/dH $>$ 100 K/T observed in Ru1212 powders suggests that
the cores of the related Josephson vertex may be as large as 10$^{-2}$-10$%
^{-1}$ $\mu $m.\cite{lor} The pinning, therefore, can be very weak if the
sample can be regarded as homogeneous over such length-scales. The fact that
full Meissner effects have been routinely observed in both underdoped and
overdoped cuprates, where evidence for a possible mesoscopic
phase-separation is accumulating,\cite{wen} supports the arguments. To
verify this, the $\Delta $M$_{FC}$/H of the nine samples at H = 5 Oe is
plotted against their 1/$\lambda ^{2}$ at 5 K (inset, Fig. 5). The rough
linear-correlation between the two parameters indicates that the Meissner
effect $\Delta $M$_{FC}$/H of Ru1222 is mainly determined by the intragrain
Josephson penetration depth $\lambda $, at least for the samples examined
here. Notice that the largest $\Delta $M$_{FC}$/H observed here (0.06 emu/cm$%
^{3}$ at 5 Oe) is already a significant fraction of 1/4$\pi $; a 100\%
Meissner effect below 1 Oe may not be sufficient evidence for
microscopic-homogeneous superconductivity.

In summary, the intragrain T$_{c}$ of Ru1222 has been tuned from 17 K to 40
K through O$_{2}$/Ar annealing. The corresponding change in the normal-state
hole-concentration, however\textit{,} is too small to account for the
change. The associated intragrain 1/$\lambda ^{2}$, on the other hand,
increases 20-fold, much more than that expected from the proposed T$_{c}$ 
\textit{vs}. 1/$\lambda ^{2}$ line for homogeneous cuprates. A
Josephson-Junction-array model, therefore, is invoked to interpret the data.
The increases in both T$_{c}$ and Meissner effect with oxygen-intake in such
a model are mainly due to the improvement of the intragrain granularity.

\begin{acknowledgments}
We thanks V. Diatschenko for careful read the manuscript.
This work is supported in part by NSF Grant, the T. L. L. Temple Foundation,
the John and Rebecca Moores Endowment and the State of Texas through TCSUH,
and in LBNL by DOE.
\end{acknowledgments}


\begin{thebibliography}{99}
\bibitem{note2} The magnetic backgound apparantly chnages with annealing. In
particular, the overall M$_{ZFC}$ is positive for sample A2. However, its
initial slope dM$_{ZFC}$(5 K)/dH (not shown here) is still negative. The
thermopower reaches zero around T$_{c}$. In addition, both the dM$_{ZFC}$/dH
and the $\Delta $M$_{ZFC}$ vary with the annealing systematically. We
therefore attribute this to a change in magnetic background.
\bibitem{note1} Some numerical coefficients at the order of unit are ignored
here.
\end{thebibliography}
\end{document}